\documentclass{ws-p10x7}

\usepackage{rotate}

\begin{document}



\thispagestyle{empty}

\onecolumn
\begin{flushright}
\begin{tabular}{l}
DESY 00--122\\
TUM--HEP--290/00\\
hep--ph/0008298\\
August 2000
\end{tabular}
\end{flushright}

\vspace*{1.5truecm}

\begin{center}
\boldmath
{\Large \bf Constraints on $\gamma$ and Strong Phases\\ 
\vspace*{0.3truecm}
from $B\to\pi K$ Decays}
\unboldmath

\vspace*{1.2cm}

{\sc{\large Andrzej J. Buras}}${}^{1,}$\footnote{E-mail: 
{\tt aburas@ally.t30.physik.tu-muenchen.de}} \,\, {\large and }\,\,
{\sc{\large Robert Fleischer}}${}^{2,}$\footnote{E-mail: 
{\tt Robert.Fleischer@desy.de}}\\[0.8cm]
\vspace*{0.1cm} ${}^1${\it Technische Universit\"at M\"unchen, 
Physik Department, D--85748 Garching, Germany}\\[0.3cm]
\vspace*{0.1cm} ${}^2${\it Deutsches Elektronen-Synchrotron DESY, 
Notkestr.\ 85, D--22607 Hamburg, Germany}

\vspace{1.3truecm}

{\large\bf Abstract\\[10pt]} \parbox[t]{\textwidth}{
As we pointed out recently, the neutral decays $B_d\to\pi^\mp K^\pm$ and 
$B_d\to\pi^0K$ may provide non-trivial bounds on the CKM angle $\gamma$. 
Here we reconsider this approach in the light of recent CLEO data, 
which look very interesting. In particular, the results for the corresponding 
CP-averaged branching ratios are in favour of strong constraints on $\gamma$, 
where the second quadrant is preferred. Such a situation would be in conflict 
with the standard analysis of the unitarity triangle. Moreover, constraints 
on a CP-conserving strong phase $\delta_{\rm n}$ are in favour of a negative
value of $\cos\delta_{\rm n}$, which would be in conflict with the 
factorization expectation. In addition, there seems to be an interesting 
discrepancy with the bounds that are implied by the charged $B\to\pi K$ 
system: whereas these decays favour a range for $\gamma$ that is similar 
to that of the neutral modes, they point towards a positive value of
$\cos\delta_{\rm c}$, which would be in conflict with the expectation of 
equal signs for $\cos\delta_{\rm n}$ and $\cos\delta_{\rm c}$.}

\vspace{1.5cm}
 
{\sl Talk given by R. Fleischer at the\\
XXXth International Conference on High Energy Physics (ICHEP 2000),\\
Osaka, Japan, 27 July -- 2 August 2000\\
To appear in the Proceedings}
\end{center}

\twocolumn

\thispagestyle{empty}

\vbox{}
\newpage
 
\addtocounter{page}{-2}


\title{Constraints on {\boldmath$\gamma$\unboldmath} and Strong 
Phases from {\boldmath$B\to\pi K$\unboldmath} Decays}

\author{Andrzej J. Buras}

\address{Technische Universit\"at M\"unchen, 
Physik Department, D--85748 Garching, Germany\\
E-mail: aburas@ally.t30.physik.tu-muenchen.de}

\author{Robert Fleischer}

\address{Deutsches Elektronen-Synchrotron DESY, 
Notkestr.\ 85, D--22607 Hamburg, Germany\\
E-mail: Robert.Fleischer@desy.de}

\twocolumn[\maketitle\abstract{

As we pointed out recently, the neutral decays $B_d\to\pi^\mp K^\pm$ and 
$B_d\to\pi^0K$ may provide non-trivial bounds on the CKM angle $\gamma$. 
Here we reconsider this approach in the light of recent CLEO data, 
which look very interesting. In particular, the results for the corresponding 
CP-averaged branching ratios are in favour of strong constraints on $\gamma$, 
where the second quadrant is preferred. Such a situation would be in conflict 
with the standard analysis of the unitarity triangle. Moreover, constraints 
on a CP-conserving strong phase $\delta_{\rm n}$ are in favour of a negative
value of $\cos\delta_{\rm n}$, which would be in conflict with the 
factorization expectation. In addition, there seems to be an interesting 
discrepancy with the bounds that are implied by the charged $B\to\pi K$ 
system: whereas these decays favour a range for $\gamma$ that is similar 
to that of the neutral modes, they point towards a positive value of
$\cos\delta_{\rm c}$, which would be in conflict with the expectation of 
equal signs for $\cos\delta_{\rm n}$ and $\cos\delta_{\rm c}$. 
}]

\section{Introduction}\label{sec:intro}
In order to obtain direct information on the angle $\gamma$ of the 
unitarity triangle of the CKM matrix, $B\to\pi K$ decays are very 
promising. In the following, we focus on our analysis Ref.\ 1, making use 
of the most recent CLEO data\cite{CLEO}. Because of the small ratio 
$|V_{us}V_{ub}^\ast/(V_{ts}V_{tb}^\ast)|\approx0.02$, $B\to\pi K$
modes are dominated by QCD penguin topologies. Due to the large 
top-quark mass, we have also to care about electroweak (EW) 
penguins. In the case of $B^0_d\to\pi^-K^+$ and $B^+\to\pi^+K^0$, 
these topologies contribute in colour-suppressed form and are hence 
expected to play a minor role, whereas they contribute in colour-allowed 
form to $B^+\to\pi^0K^+$ and $B^0_d\to\pi^0K^0$ and may here even compete 
with tree-diagram-like topologies.

So far, strategies to probe $\gamma$ through $B\to\pi K$ decays have 
focused on the following two systems: $B_d\to\pi^\mp K^\pm$, 
$B^\pm\to\pi^\pm K$ (``mixed'')\cite{BpiK-mixed,FM}, and $B^\pm\to\pi^0K^\pm$, 
$B^\pm\to\pi^\pm K$ (``charged'')\cite{NR}. Recently, we pointed 
out that also the neutral combination $B_d\to\pi^\mp K^\pm$,
$B_d\to\pi^0 K$ is very promising\cite{BF-BpiK}.

\section{Constraints on \boldmath$\gamma$\unboldmath }\label{sec:gamma}
Interestingly, already CP-averaged branching ratios may lead to 
highly non-trivial constraints on $\gamma$. Here the key quantities 
are 
\begin{eqnarray}
R&\equiv&\frac{\mbox{BR}(B_d\to\pi^\mp K^\pm)}{\mbox{BR}(B^\pm\to\pi^\pm K)}
=0.95\pm0.28\label{CLEO-mix}\\
R_{\rm c}&\equiv&
\frac{2\mbox{BR}(B^\pm\to\pi^0K^\pm)}{\mbox{BR}(B^\pm\to\pi^\pm K)}
=1.27\pm0.47~~\mbox{}\label{CLEO-charged}\\
R_{\rm n}&\equiv&
\frac{\mbox{BR}(B_d\to\pi^\mp K^\pm)}{2\mbox{BR}(B_d\to\pi^0 K)}
=0.59\pm0.27,\label{CLEO-neut}
\end{eqnarray}
where we have also taken into account the CLEO results reported in 
Ref.\ 2. If we employ the $SU(2)$ flavour symmetry and
certain dynamical assumptions, concerning mainly the smallness of
FSI effects, we may derive a general parametrization\cite{BF-BpiK} for
(\ref{CLEO-mix})--(\ref{CLEO-neut}),
\begin{equation}
R_{({\rm c,n})}=R_{({\rm c,n})}(\gamma,q_{({\rm c,n})},r_{({\rm c,n})},
\delta_{({\rm c,n})}),
\end{equation}
where $q_{({\rm c,n})}$ denotes the ratio of EW penguins to ``trees'', 
$r_{({\rm c,n})}$ is the ratio of ``trees'' to QCD penguins, and 
$\delta_{({\rm c,n})}$ is the CP-conserving strong phase between 
``tree'' and QCD penguin amplitudes. The parameters $q_{({\rm c,n})}$
can be fixed through theoretical arguments: in the ``mixed'' system,
we have $q\approx0$, as EW penguins contribute only in colour-suppressed
form; in the charged\cite{NR} and neutral\cite{BF-BpiK} $B\to\pi K$ systems,
$q_{\rm c}$ and $q_{\rm n}$ can be fixed through the $SU(3)$ flavour
symmetry without dynamical assumptions. The $r_{({\rm c,n})}$ can
be determined with the help of additional experimental information:
in the ``mixed'' system, $r$ can be fixed through arguments based on
``factorization'', whereas $r_{\rm c}$ and $r_{\rm n}$ can be determined
from $B^+\to\pi^+\pi^0$ by using only the $SU(3)$ flavour symmetry. 

At this point, a comment on FSI effects is in order. Whereas the 
determination of $q$ and $r$ as sketched above may be affected by FSI 
effects, this is {\it not} the case for $q_{\rm c,n}$ and $r_{\rm c,n}$, 
since here $SU(3)$ suffices. Nevertheless, we have to assume that 
$B^+\to\pi^+K^0$ or $B_d^0\to\pi^0K^0$ do {\it not} involve a CP-violating 
weak phase:
\begin{eqnarray}
\lefteqn{A(B^+\to\pi^+K^0)=-\,|\tilde P|e^{i\delta_{\tilde P}}}\nonumber\\
&&\qquad=A(B^-\to\pi^-\overline{K^0}).
\end{eqnarray}
This relation may be affected by rescattering processes such as 
$B^+\to\{\pi^0K^+\}\to\pi^+K^0$:
\begin{displaymath}
A(B^+\to\pi^+K^0)=-\,|\tilde P|e^{i\delta_{\tilde P}}\left[1+
\rho_{\rm c}\, e^{i\theta}e^{i\gamma}\right],
\end{displaymath}
where $\rho_{\rm c}$ is doubly Cabibbo-suppressed and is naively expected to
be negligibly small. In the ``QCD factorization'' approach\cite{BBNS}, 
there is no significant enhancement of $\rho_{\rm c}$ through rescattering 
processes. However, there is still no theoretical consensus on the importance 
of FSI effects. In the charged $B\to\pi K$ strategy to probe $\gamma$, they 
can be taken into account through $SU(3)$ flavour-symmetry arguments and 
additional data on $B^\pm\to K^\pm K$ decays. The present experimental upper 
bounds on these modes are not in favour of dramatic effects. In the case of 
the neutral strategy, FSI effects can be included in an {\it exact manner} 
with the help of the mixing-induced CP asymmetry 
${\cal A}_{\rm CP}^{\rm mix}(B_d\to\pi^0K_{\rm S})$\cite{BF-BpiK}.

\begin{figure}
\epsfxsize160pt
\rotate[r]{
\figurebox{}{}{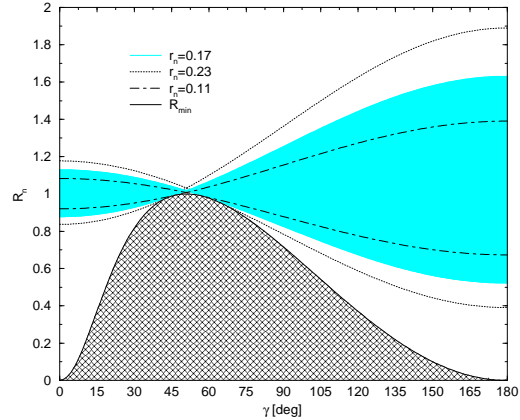}}
\caption{The dependence of the extremal values of $R_{\rm n}$ (neutral
$B\to\pi K$ system) 
on $\gamma$ for $q_{\rm n}=0.63$.}\label{fig1}
\vspace*{-0.5truecm}
\end{figure}

\begin{figure}
\epsfxsize108pt
\vspace*{-0truecm}
\rotate[r]{
\figurebox{}{}{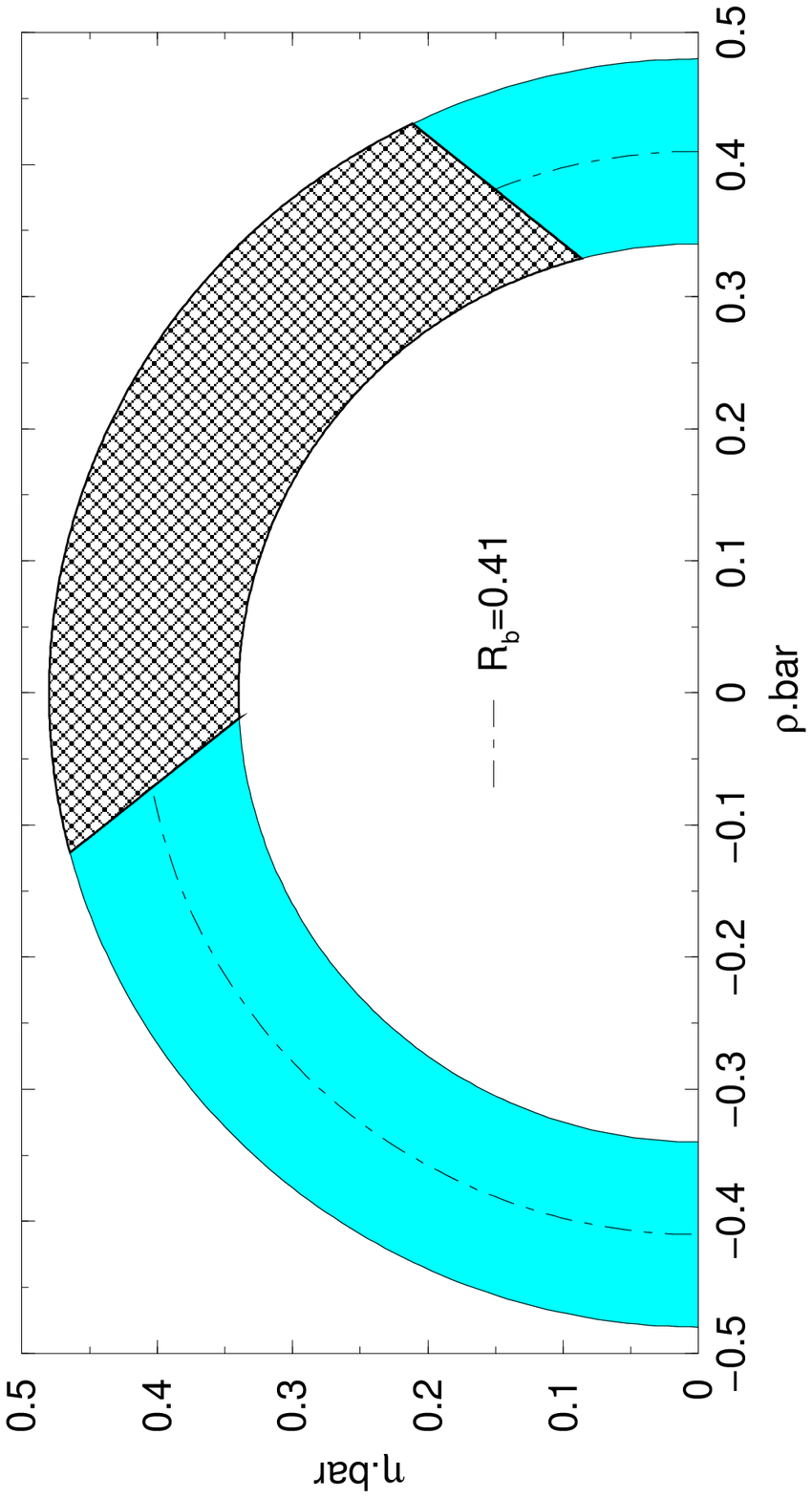}}
\caption{The constraints in the $\overline{\varrho}$--$\overline{\eta}$ 
plane implied by (\ref{const2}) for $R_{\rm n}=0.6$
and $q_{\rm n}=0.63 \times[0.41/R_b]$.}\label{fig2}
\vspace*{-0.5truecm}
\end{figure}

In contrast to $q_{({\rm c,n})}$ and $r_{({\rm c,n})}$, the strong phase 
$\delta_{({\rm c,n})}$ suffers from large hadronic uncertainties and is 
essentially unknown. However, we can get rid of $\delta_{({\rm c,n})}$ 
by keeping it as a ``free'' variable, yielding minimal and maximal values 
for $R_{({\rm c,n})}$:
\begin{equation}\label{const1}
\left.R^{\rm ext}_{({\rm c,n})}\right|_{\delta_{({\rm c,n})}}=
\mbox{function}(\gamma,q_{({\rm c,n})},r_{({\rm c,n})}).
\end{equation}
Keeping in addition $r_{({\rm c,n})}$ as a free variable, we obtain 
another -- less restrictive -- minimal value for $R_{({\rm c,n})}$:
\begin{equation}\label{const2}
\left.R^{\rm min}_{({\rm c,n})}\right|_{r_{({\rm c,n})},\delta_{({\rm c,n})}}
=\kappa(\gamma,q_{({\rm c,n})})\sin^2\gamma.
\end{equation}
In Fig.\ \ref{fig1}, we show the dependence of (\ref{const1}) and
(\ref{const2}) on $\gamma$ for the neutral $B\to\pi K$ system\footnote{The
charged $B\to\pi K$ curves look very similar.}.
Here the crossed region below the $R_{\rm min}$ curve, which is described
by (\ref{const2}), is excluded. On the other hand, the shaded region 
is the allowed range (\ref{const1}) for $R_{\rm n}$, arising 
in the case of $r_{\rm n}=0.17$. Fig.~\ref{fig1} allows us to read off 
immediately the allowed region for $\gamma$ for a given value 
of $R_{\rm n}$. Using the central value of the present CLEO result 
(\ref{CLEO-neut}), $R_{\rm n}=0.6$, the $R_{\rm min}$ curve implies 
$0^\circ\leq\gamma\leq21^\circ\,\lor\,100^\circ\leq\gamma
\leq180^\circ$. The corresponding situation in the 
$\overline{\varrho}$--$\overline{\eta}$ plane is shown in Fig.~\ref{fig2},
where the crossed region is excluded and the circles correspond to
$R_b=0.41\pm0.07$. As the theoretical expression for $q_{\rm n}$ is 
proportional to $1/R_b$, the constraints in the 
$\overline{\varrho}$--$\overline{\eta}$ plane are actually more appropriate
than the constraints on $\gamma$.

If we use additional information on the parameter $r_{\rm n}$, we may 
put even stronger constraints on $\gamma$. For $r_{\rm n}=0.17$, we obtain, 
for instance, the allowed range $138^\circ\leq\gamma\leq180^\circ$. 
It is interesting to note that the
$R_{\rm min}$ curve is only effective for $R_{\rm n}<1$, which is
favoured by the most recent CLEO data\cite{CLEO}. A similar pattern
is also exhibited by the first BELLE results\cite{BELLE} presented at this 
conference, yielding $R_{\rm n}=0.4\pm0.2$. 

For the central value $R_{\rm c}=1.3$ of (\ref{CLEO-charged}), 
(\ref{const2}) is not effective and $r_{\rm c}$ has to be fixed in order to 
constrain $\gamma$. Using $r_{\rm c}=0.21$, we obtain 
$87^\circ\leq\gamma\leq180^\circ$. Although it is too early to draw definite 
conclusions, it is important to emphasize that the most recent CLEO results 
on $R_{({\rm c,n})}$ prefer the second quadrant for $\gamma$, 
i.e.\ $\gamma\geq 90^\circ$. Similar 
conclusions were also obtained using other $B\to\pi K$, $\pi\pi$ 
strategies\cite{Hou}. Interestingly, such a situation would be in conflict 
with the standard analysis of the unitarity triangle\cite{AL}, yielding 
$38^\circ\leq\gamma\leq81^\circ$.

\section{Constraints on Strong Phases}\label{sec:strong}
The $R_{({\rm c,n})}$ allow us to determine $\cos\delta_{({\rm c,n})}$ as 
functions of $\gamma$, thereby providing also constraints on the strong 
phases $\delta_{({\rm c,n})}$\cite{BF-neut}.
Interestingly, the present CLEO data are in favour of $\cos\delta_{\rm n}<0$, 
which would be in conflict with ``factorization''. Moreover, they point 
towards a positive value of $\cos\delta_{\rm c}$, which would be in conflict 
with the theoretical expectation of equal signs for $\cos\delta_{\rm c}$ and 
$\cos\delta_{\rm n}$. 

\section{Conclusions and Outlook}\label{sec:concl}
If future data should confirm the ``puzzling'' situation for $\gamma$
and $\cos\delta_{{\rm c,n}}$ favoured by the present $B\to\pi K$ CLEO 
data, it may be an indication for new-physics contributions to the EW penguin 
sector, or a manifestation of flavour-symmetry-breaking effects. In order 
to distinguish between these possibilities, further studies are needed. 
As soon as CP asymmetries in $B_d\to\pi^\mp K^\pm$ or $B^\pm\to\pi^0K^\pm$ 
are observed, we may even {\it determine} $\gamma$ and $\delta_{({\rm c,n})}$. 
Here we may also arrive at a situation, where the $B\to\pi K$ observables 
do not provide any solution for these quatities\cite{FMat}, which would be an 
immediate indication for new physics. We look forward to new data from 
the $B$-factories.

\end{document}